\documentstyle[12pt]{article}
\begin{document}
\newcommand{\be}{\begin{equation}}
\newcommand{\ee}{\end{equation}}
\renewcommand{\thefootnote}{\fnsymbol{footnote}}
\begin{flushright}
CERN-TH.6914/93

\noindent
POLFIS-TH.04/93

\noindent

UCLA/93/TEP/18
\noindent
\end{flushright}
\vskip 35pt
\begin{center}{\Large\bf Five-Brane Effective Field Theory on Calabi-Yau
Threefolds}
\end{center}
\vskip .5in
\begin{center}{
{\large R. D'Auria}
\footnote{Work supported in part by the United States Department of Energy,
under contract No. DOE-ATO3-88ER40384, Task C, by the National Science
Foundation under grant No. PHY89-04035; and by M.U.R.S.T.}\\
Dipartimento di Fisica, Politecnico di Torino
and INFN, Sezione di Torino, Italy\\
\vskip .1in
{\large S. Ferrara}\footnotemark[1]\\
Theory Division, CERN, CH1211, Geneva 23, Switzerland\\
\vskip .1in
and\\
\vskip .1in
{\large M. Villasante}\footnote
{Work supported in part by the World Laboratory.}\\
Department of Physics, University of California, Los Angeles, CA 90024-1547}

\end{center}
\vskip 1in
\begin{center}{\bf\large Abstract}
\end{center}
\vskip 12pt

	We consider the compactification of the dual form of $N=1$ $D=10$
supergravity on a six-dimensional Calabi-Yau manifold. An $N=1$ off-shell
supergravity effective Lagrangian in four dimensions can be constructed
in a dual version of the gravitational sector (new-minimal supergravity
form). Superspace duality has a simple interpretation in terms of Poincar\'{e}
duality of two-form cohomology.

	The resulting $4D$ Lagrangian may describe the low-energy point-field
limit of a five-brane theory, dual to string theory, provided Calabi-Yau
spaces are consistent vacua of such dual theory.
\vskip 30pt

\noindent
CERN-TH.6914/93

\noindent
June 1993
\thispagestyle{empty}
\addtocounter{page}{-1}
\newpage

\thispagestyle{plain}
\hoffset = 0.5in
\textwidth 6.5in
\textheight 9in
\renewcommand{\theequation}{\arabic{section}.\arabic{equation}}
\renewcommand{\baselinestretch}{1.7}

\noindent
\section{Introduction}

In recent years it has been shown that five-branes arise as ``soliton"
solutions \cite{uno} of the effective theory describing $10D$ heterotic strings
\cite{dos},
i. e. $N=1$ ten dimensional supergravity with a $2$-form coupled to
anomaly-free
Yang-Mills multiplets \cite{tres}.
In this respect, five-branes solutions are the
analog of the monopole solutions of $4D$ gauge theories and they are
non-perturbative in nature \cite{cuatro,cinco}.

This allowed to speculate that if a fundamental theory of five-branes
could be found , this theory would probably exchange weak and strong coupling
regimes of string theory \cite{seis,siete},
by allowing us to get non-perturbative results of
string dynamics by perturbative calculations of five-brane interactions.

In the case of a five-brane, the point-field theory limit in $D=10$ is $N=1$
supergravity in a dual formulation, i. e. with a six-form antisymmetric tensor
[8-11].

The aim of this paper is to present the most general $N=1$ off-shell
supergravity
action of a five-brane in D=4 with six dimensions compactified on a Calabi-Yau
manifold \cite{once}.
Five-branes compactified on a six-dimensional torus have already
been investigated and there \cite{doce},
the powerful constraints of $N=4$ supersymmetry have
been extensively used to discuss the symmetry properties of the massive
excitations \cite{doce,trece}.

Here we consider the effective theory in a case of $N=1$ unbroken supergravity
in D=4 where it is also possible to present an off-shell formulation of the
theory.

\vskip .1in
The main results of this investigation are:

\vskip .1in
\noindent
1) The classical low-energy effective theory of $10D$ five-brane compactified
on
a Calabi-Yau manifold can not be written off-shell in standard supergravity
form but a ``dual" version of auxiliary supergravity fields must be used
\cite{ventinueve}.

\vskip .1in
\noindent
2) Linear multiplets \cite{catorce},
describing the zero modes of the Calabi-Yau manifold
are naturally associated to the $H^{(2,2)}$ cohomology which is dual to the
$H^{(1,1)}$ cohomology. This is encoded in a change of coordinates in the
moduli space that has an obvious superspace interpretation.

\vskip .1in
\noindent
3) Mirror symmetry can not be imposed any longer on the tree-level Lagrangian
and can only arise if the space-time manifold $M_{4}$ is equipped with a
non-trivial topology \cite{quince}.
This statement supports the folklore that tree-level
results of string theory should arise as non-perturbative effects in the dual
theory \cite{cuatro,cinco,dieciseis}. It also confirms some evidence that
the $\sigma$-model and loop expansions are ``dual" in the five-brane
formulation
of the theory.

\vskip .1in
\noindent
4) The dilaton chiral multiplet is no longer protected by a perturbative
Peccei-Quinn symmetry and its axion may have a discrete shift symmetry.

\newpage

\noindent
\section{Four-dimensional reduction of ten dimensional
 supergravity with a $6$-form and the zero mode
 analysis.}

We would like to consider in this section some properties of the dual
$10D$ supergravity theory when a six-form (rather than a two-form) is
introduced in the theory [8-11].

Let us confine our analysis to the bosonic massless degrees of freedom
of the gravitational sector, including the antisymmetric tensor as well
as the dilaton. Let us first recall the zero mode analysis for the
heterotic superstring theory.
The bosonic massless fields of heterotic string theory are: the graviton
$G_{\hat{\mu} \hat{\nu}}$, the $2$-form
$B_{\hat{\mu} \hat{\nu}}$, $\hat{\mu}, \hat{\nu} =
1,..., 10$ and the dilaton $\Phi$. After compactification of six dimensions
on a Calabi-Yau manifold they give rise to the following massless fields
in four dimensions

\[ G_{\hat{\mu} \hat{\nu}} \rightarrow (g_{\mu \nu}, g_{i \,
\overline{\jmath}},
g_{i j}, g_{\overline{\imath} \overline{\jmath}}) \]

\be
\hat{\mu} = (\mu, i, \overline{\imath}) \;\; \mu = 1,..,4 \;\;
i, \overline{\imath} = 1, 2, 3
\ee
\vskip .1in

\noindent
where $g_{\mu \nu}$ is the graviton field, and the internal components of
the metric give

\be
g_{i \, \overline{\jmath}} \rightarrow (0; (1, 1))- form \rightarrow h_{(1,1)}
\;\; real \;\; scalars
\ee

\be
g_{i j} \rightarrow (0; (2, 1))-form \rightarrow h_{(2,1)}
\;\; complex \;\; scalars
\ee

\vskip .1in
\noindent
where we have denoted in brackets the degree of forms in $M_{4} \times K_{6}$.
For the ten-dimensional $2$-form we  get

\be
B_{\hat{\mu} \hat{\nu}} \rightarrow (b_{\mu \nu}, b_{i \, \overline{\jmath}})
\ee

\noindent
where $b_{\mu \nu}$ is a $(2;(0,0)$-form,dual to a pseudoscalar in space-time
, while the internal components $b_{i\bar j}$ give rise to space-time
scalars:

\be
b_{i \, \overline{\jmath}} \rightarrow (0; (1, 1))- form
\rightarrow h_{(1,1)} \;\; real
\;\; scalars
\ee

\noindent
Finally $\Phi$ gives rise to a $(0;(0,0))$-form, i.e.the 4-dimensional dilaton:
\be
\Phi \rightarrow dilaton \;\; (real \;\; scalar)
\ee
\vskip .1in

The supersymmetric multiplets are: the gravity multiplet (graviton $g_{\mu
\nu}$
$\oplus$ gravitino $\psi_{\mu}$); the linear multiplet [19-22]
($\Phi$, $b_{\mu \nu}$
$\oplus$ dilatino $\chi$); $h_{(1,1)}$ chiral multiplets
($g_{i \, \overline{\jmath}}$, $b_{i \, \overline{\jmath}}$ $\oplus$ fermionic
partners); and $h_{(2,1)}$ chiral multiplets ($g_{i j}$ $\oplus$ fermionic
partners).

There is a (string) perturbative Peccei-Quinn symmetry due to the $b_{\mu \nu}$
gauge symmetry, a perturbative $\sigma$-model Peccei-Quinn symmetry on the
$b_{i \, \overline{\jmath}}$, the last being broken by world-sheet
non-perturbative effects \cite{diecisiete,PCXO}, that is when

\be
\int b_{i \, \overline{\jmath}} dx^{i}d\overline{x}^{\overline{\jmath}} \neq 0
\ee
\vskip .1in

\noindent
corresponding to the presence of world-sheet instantons.

The $b_{\mu \nu}$ gauge symmetry can be broken when \cite{quince}

\be
\int H_{\mu \nu \rho} dx^{\mu} dx^{\nu} dx^{\rho} \neq 0
\ee

\noindent
i.e. if $H^{3}(M_{4},\bf Z) \neq 0$ and $b_{\mu \nu}$ becomes a singular gauge
field.

In the dual $10D$ supergravity theory, $B_{\hat{\mu} \hat{\nu}}$ is replaced by
$A_{\hat{\mu}_{1}...\hat{\mu}_{6}}$. There are the same number of massless
modes but now (2.5-2.7) are replaced by

\be
A_{\hat{\mu}_{1}...\hat{\mu}_{6}} \rightarrow (b \; \epsilon_{i j k}
\overline{\epsilon}_{\overline{\imath} \overline{\jmath} \overline{k}},
b_{\mu \nu i \, \overline{\jmath} l \overline{k}})
\ee

\noindent
where
\be
b \epsilon_{i j k} \overline{\epsilon}_{\overline{\imath} \overline{\jmath}
\overline{k}} \rightarrow (0;(3, 3))- form \rightarrow 1 \; pseudoscalar
\ee

\be
b_{\mu \nu i \, \overline{\jmath} l \overline{k}} \rightarrow  (2; (2,2))- form
\rightarrow  h_{(1,1)}\;\; antisymmetric \;\; tensors.
\ee

\vskip .1in
So in the supergravity sector the multiplets are now: the dilaton chiral
multiplet ($\Phi$, $b$ $\oplus$ fermionic partners) and $h_{(1,1)}$ linear
multiplets ($g_{i \, \overline{\jmath}}$,
$b_{\mu \nu i \, \overline{\jmath} l \overline{k}}$, $\oplus$ fermionic
partners). The other multiplets remain unchanged.

We observe that the linear multiplets are given in terms of
$(1,1)$ and $(2,2)$ forms so it is
natural to
replace $g_{i \, \overline{\jmath}}$ with its dual form defined as follows:

\be
 ^{\ast}J = g_{i \, \overline{\jmath} l \overline{k}}
dx^{i} d\overline{x}^{\overline{\jmath}} dx^{l} d\overline{x}^{\overline{k}}
\ee

\vskip .1in
\noindent
where $^{\ast}J$ is the dual of the K\"{a}hler form $J = i g_{i \,
\overline{\jmath}} dx^{i} d\overline{x}^{\overline{\jmath}}$.

In the five-brane theory, the perturbative and non-perturbative symmetries
are best seen by looking to the $p$-brane sheet Wess-Zumino term:

\be
(1-brane \equiv String) \rightarrow \int B_{\hat{\mu} \hat{\nu}} dx^{\hat{\mu}}
dx^{\hat{\nu}}
\ee

\be
(5-brane) \rightarrow \int A_{\hat{\mu_{1}}... \hat{\mu_{6}}}
dx^{\hat{\mu_{1}}}...dx^{\hat{\mu_{6}}}
\ee

\vskip .1in
As a consequence of (2.10) the background fields which contribute to eq.
(2.15) are

\be
\int b V_{6}
\ee

\be
\int b_{\mu \nu i \, \overline{\jmath} l \overline{k}} dx^{\mu}dx^{\nu}
dx^{i}d\overline{x}^{\overline{\jmath}}dx^{l}d\overline{x}^{\overline{k}}
\ee

\vskip .1in
\noindent
where $V_{6}$ denotes the volume form on the Calabi-Yau manifold.

The Peccei-Quinn symmetry on $b$ is broken by classical
effects since we may have \cite{trece}

\be
\int b V_{6} \neq 0
\ee

\vskip .1in
\noindent
while the Peccei-Quinn symmetry for the $(2,2)$ forms can be broken if the
gauge invariance of the $b_{\mu \nu A}$, $A = 1,\dots,h_{(1,1)}$,
$2$-forms is broken. This
can only happen if space-time has a non-trivial topology as was the case for
the
dilaton multiplet (see eq. (2.9)) in the dual string theory \cite{quince}.

Comparing eqs. (2.16) and (2.17) to eqs. (2.7) and (2.8) we see that the roles
of the dilaton and moduli multiplets have been interchanged, supporting the
idea that what is perturbative in strings is non-perturbative in five-branes
and viceversa \cite{uno,cinco,doce,trece,ventisiete}.

\vskip 30pt
\noindent
\section{Supersymmetry and duality transformations.}
\setcounter{equation}{0}

The string-effective theory for the $(1,1)$ and $(2,1)$ moduli and dilaton
sector is given by the following superfield expression for the Lagrangian
density [25-30]

\be
{\cal L} =
\Phi(T^{A} + \overline{T}^{A}, \psi_{\alpha}, S) S_{0} \overline{S}_{0}
\mid_{D}.
\ee

\vskip .1in
Here $T^{A}$, $\psi_{\alpha}$ are the $(1,1)$ and $(2,1)$ moduli, $S$ is the
(dual) dilaton multiplet and $\Phi$ is defined by

\be
\Phi = \left[ d_{A B C} (T + \overline{T})^{A}
(T + \overline{T})^{B} (T + \overline{T})^{C} \right]^{1/3}
(S + \overline{S})^{1/3} e^{- K_{2}(\psi)/3}.
\ee

\vskip .1in
\noindent
$d_{A B C}$ are the intersection numbers  of the Calabi-Yau manifold
and the K\"{a}hler potential $K$, which is additive in the three sectors of
fields $T^{A}$, $\psi$ and $S$ is given by
\cite{venticuatro}

\be
K = - 3 ln \Phi
\ee

\noindent
$S_{0}$ is the chiral compensator of conformal supergravity [31-33],
which after
superconformal gauge fixing, gives the off-shell standard
supergravity multiplet

\be
(e_{a \mu}, \psi_{\mu}, A_{\mu}, S, P)
\ee

\vskip .1in
To go to the five-brane formulation, one should perform a duality
transformation
\cite{ventiseis}
between the $T^{A}$ multiplets and the $h_{(1,1)}$ linear multiplets $L_{A}$.
However, this is impossible for $\Phi$ homogenous of degree $1$ in the $T^{A}$
fields. Indeed, the duality transformation, which amounts to replace
$Re \, T^{A}$ by the real linear multiplet $L_{A}$ via a superspace
Legendre transformation \cite{ventiseis}, would give

\be
\Phi_{A} = L_{A}
\ee

\vskip .1in
\noindent
where $\Phi_{A} = \frac{\partial \Phi}{\partial T^{A}} =
 \frac{\partial \Phi}{\partial ( T^{A}+\bar T^{A})}$ . This implies

\be
\frac{\partial \Phi_{A}}{\partial L_{B}} =
\frac{\partial (T + \overline{T})^{C}}{\partial L_{B}} \Phi_{A C} =
\delta_{A}^{B}
\ee

\noindent
which gives

\be
\frac{\partial (T + \overline{T})^{C}}{\partial L_{B}} \Phi_{A C}
(T + \overline{T})^{A} =
(T + \overline{T})^{B}
\ee

\vskip .1in
\noindent
and this is impossible since $\Phi_{A C} (T + \overline{T})^{A} = 0$.
The lack of invertibility
of this relation was noticed in \cite{treintaytres} in the context of no-scale
supergravity models \cite{treintaycuatro}. Recently
it has been discussed in \cite{ventisiete} for
an orbifold compactification of the $10D$ five-brane.

This difficulty means that linear multiplets with particular
couplings cannot be coupled
to standard supergravity.

However we can cure this difficulty by going to the dual formulation of
supergravity by ``dualizing" \cite{ventiseis}
the compensator $S_{0}$ into a linear multiplet
compensator $L_{0}$.

This is done by replacing $S_{0} \overline{S}_{0} \rightarrow e^{U}$ and by
imposing the
condition $U = \Sigma + \overline{\Sigma}\ ( S_0 =e^{\Sigma})$ by adding the
Lagrange multiplier term
$-U L_{0}$ to the action (3.1).

A trivial calculation gives for the dual form of eq. (3.1)

\be
\tilde{\cal L} = - L_{0} ln L_{0} + L_{0} ln \Phi\;=\;L_0\;ln{\Phi/L_0}
\ee

\vskip .1in
\noindent
which describes a matter system
coupled to ``new-minimal" supergravity with a ``dual" gravity
multiplet given by \cite{ventinueve,sabi}

\be
(e_{a \mu}, \psi_{\mu}, A_{\mu}, a_{\mu \nu})
\ee

\vskip .1in
Since $ln \Phi$ is additive in the different multiplets we may now dualize
separately the term in $\Phi$ containing the
$T^{A}$ scalars:

\be
\frac{1}{3} \, L_{0} \; ln\left[ d_{A B C} (T + \overline{T})^{A} (T +
\overline{T})^{B}
(T + \overline{T})^{C} \right]
\ee

\vskip .1in
\noindent
which must be replaced by

\be
\frac{1}{3} L_{0} ln (d_{A B C} U^{A} U^{B} U^{C}) - U^{A} L_{A}
\ee

\vskip .1in
It is immediate to see that eq. (3.11) gives rise to the following term in
the modified Lagrangian
of $\tilde{\cal L}$

\be
\frac{1}{3} \, L_{0} \; ln\left[ d_{A B C} (T + \overline{T})^{A}(L/L_{0})
(T + \overline{T})^{B}(L/L_{0}) (T + \overline{T})^{C}(L/L_{0}) \right]
\ee

\vskip .1in
\noindent
where $(T + \overline{T})^{A}(L/L_{0})$ is the solution of the equations

\be
\frac{d_{A B C} (T + \overline{T})^{B}
(T + \overline{T})^{C}}
{d_{D E F} (T + \overline{T})^{D} (T + \overline{T})^{E}
(T + \overline{T})^{F}}
= \frac{L_{A}}{L_{0}}
\ee

\vskip .1in
\noindent
Eq. (3.13) can be inverted. Indeed, if one takes

\be
G_{A B} = - \partial_{A} \partial_{B} ln(\Phi^{3})
\;\;\;\; \left(\partial_{A} = \frac{\partial}{\partial T^{A}} \right)
\ee

\vskip .1in
\noindent
one clearly realizes that expression (3.13) can be rewritten as

\be
\frac{1}{3} G_{A B} (T + \overline{T})^{B} = \frac{L_{A}}{L_{0}}
\ee

\vskip .1in
\noindent
with $G_{A B}$ invertible.

\noindent
The final Lagrangian is therefore

\be
{\cal L_{TOTAL}}= \;-L_{0}\;ln\;L_{0}\;+\;\widetilde{\widetilde{\cal L}}
\ee

where

\be
\widetilde{\widetilde{\cal L}} =
L_{0} \left[
{\cal F} \left( \frac{L_{A}}{L_{0}} \right) +
\frac{1}{3} ln(S + \overline{S}) - \frac{1}{3} K_{2}(\psi) \right]
\mid_{D}
\ee

\vskip .1in
\noindent
with ${\cal F} \left( L_{A}/L_{0} \right)$ given by (3.12) without
the factor $L_{0}$.
\noindent

To make the dependence in ${\cal F}$ explicit it is convenient to change
variables in eq.(3.11)

\be
U^{A}\;=\;{\hat U}^{A}\;R\;\;\;\;\;\;\;\;\;R\;=\;\left(d_{ABC} U^{A}U^{B}U^{C}
\right)^{1/3}
\ee

\noindent
Then eq.(3.13) becomes

\be
R\;=\; \frac{L^{0}}{L_{A}\hat U^{A}}\;\;\;\;\;\;\left(d_{ABC} {\hat U}^A
{\hat U}^B {\hat U}^C \;=1\right)
\ee

\noindent
with ${\hat U}^A$ homogeneous (of degree $0$) in the $L_A$'s given by

\be
d_{ABC}{\hat U}^B{\hat U}^C{\hat U}^D \; L_D\;=L_A
\ee
\noindent
The function $\cal F$ becomes
\be
{\cal F} \left(L_A/L_0\right)=ln R=ln\frac{ L_0}{L_A {\hat U}^A}
\ee

The linear multiplet contribution in ${\cal L}_{TOTAL}$ is therefore given by

\be
-L_0ln L_A{\hat U}^A(L)
\ee

\noindent
where

\be
L_A \frac{\partial}{\partial L_A}{\hat U}^B (L)=0.
\ee
Hence the Lagrangian (3.16) cannot be written in standard supergravity form
because the inverse duality transformation $L_{0} \rightarrow S_{0}$ can not
be performed due to the fact that $L_{0}$ appears linearly in eq. (3.16).

We remark that this action at the quantum level is potentially anomalous
due to the gauge invariance implied by the auxiliary field sector of the
supergravity multiplet \cite{bau,treintap}. If therefore a consistent quantum
theory exists, a quantum anomaly cancellation should take place.

\vskip 30pt

\noindent
\section{Results at the component level.}
\setcounter{equation}{0}

The most general $N=1$ supergravity Lagrangian in new-minimal formulation
containing at most two derivatives is described by 3 functions: a real function
$F(S_{i}, \overline{S_{i}}, L_{A})$ of a set of matter (chiral and linear)
multiplets, a holomorphic function $f(S_{i})$ of the chiral multiplets and
a Yang-Mills covariant symmetric holomorphic tensor $f_{a b}(S_{i})$. The
Poincar\'{e} linear multiplets $L_{A}$ correspond to the ratios
$L_{A}/L_{0}$ of conformal multiplets of the previous section, while
the Poincar\'{e} function $F$ is related to the conformal function
$\cal F$ after superconformal gauge fixing by $F = - 3 {\cal F}$. The
bosonic part of the Lagrangian has been given in \cite{sabi}:

\begin{eqnarray}
e^{-1} {\cal L}_{BOS} & = & \frac{1}{2} (1 - (nz)^{j} F_{j}) (R + 6
H_{\mu}H^{\mu}) +
(2 A^{-}_{\mu} + i F_{j} D^{-}_{\mu} z^{j} - i
F_{\overline{\jmath}} D^{-}_{\mu} \, \overline{z}^{\overline{\jmath}}) H^{\mu}
\nonumber \\
 & &+ \frac{1}{4} F^{A B} \partial_{\mu} C_{A} \partial^{\mu} C_{B}
 - \frac{1}{4} F^{A B} {\cal V}_{\mu A} {\cal V}^{\mu}_{B} \nonumber \\
 & & + \frac{i}{2} ( F^{A} \, _{j} D^{-}_{\mu} z^{j} -
F^{A} \, _{\overline{\jmath}} D^{-}_{\mu} \, \overline{z}^{\overline{\jmath}})
{\cal V}^{\mu}_{A} \nonumber \\
 & & - F_{i \overline{\jmath}} D^{-}_{\mu} z^{i} D^{- \mu} \,
\overline{z}^{\overline{\jmath}}
+ F_{i \overline{\jmath}} h^{i} \overline{h}^{\overline{\jmath}}
+ f_{j} h^{j} + \overline{f}_{\overline{\jmath}}
\overline{h}^{\overline{\jmath}} \nonumber \\
 & & - \frac{1}{4} Re \, f_{a b} \, F^{a}_{\mu \nu} F^{b \, \mu \nu}
 + \frac{1}{4} Im \, f_{a b} \, F^{a}_{\mu \nu} \, ^{\ast}F^{b \, \mu \nu}
\nonumber \\
 & & + \frac{1}{2} Re \, f_{a b} \, D^{a} D^{b}
+ ({\cal T}_{a} z)^{j} F_{j} D^{a} \;\; ,
\end{eqnarray}

\[
{\cal V}_{A}^{\mu} = - \frac{1}{2} e^{-1} \epsilon^{\mu \nu \rho \sigma}
\partial_{\nu} b_{\rho \sigma A} - 2 C_{A} H^{\mu}
\]

\noindent
where $H^{\mu}$ is the field strength of the auxiliary field $a_{\mu \nu}$
in (3.9), $A^{-}_{\mu} = A_{\mu}- 3 H_{\mu}$
and $D^{-}_{\mu}$ are covariant derivatives with a modified chiral
connection (see \cite{sabi}). In eq. (4.1) upper indices on $F$ denote
differentiation with respect to the linear multiplet scalars ($C_{A}$) and
lower indices differentiation with respect to the chiral and antichiral
ones ($z^{i}$, $\overline{z}^{\overline{\imath}}$).

We start with a function $F((T + \overline{T})^{A}, Z_{i})$ that does not
contain linear multiplets but where some of the chiral ones ($T^{A}$) appear
only in the particular combination $(T + \overline{T})^{A}$. These multiplets
can appear in $f_{a b}$ in the same real combination (though we will not
consider it here), but must be absent from
the superpotential $f$. The terms of
the bosonic Lagrangian (4.1) containing fields of these multiplets are
\begin{eqnarray}
e^{-1} \hat{\cal L} & = & i(F_{A} \partial_{\mu} t^{A} -
F_{A} \partial_{\mu} \overline{t}^{A}) H^{\mu} -
F_{A B} \partial_{\mu} t^{A} \partial^{\mu} t^{B} -
F_{A \overline{\jmath}} \partial_{\mu} t^{A} D^{- \mu} \, \overline{z}^{
\overline{\jmath}} \nonumber \\
 & & - F_{j A} D^{-}_{\mu} z^{j} \partial^{\mu} \overline{t}^{A}
+ F_{A B} h^{A} \overline{h}^{B} +
F_{A \overline{\jmath}} h^{A} \overline{h}^{\overline{\jmath}} +
F_{j B} h^{j} \overline{h}^{B}
\end{eqnarray}

Since we intend to dualize the Lagrangian in (4.2) into one containing linear
multiplets instead of the chiral $T^{A}$, we must eliminate the auxiliary
fields $h^{A}$ from their equations of motion:
\begin{eqnarray}
h^{A} = - (F^{-1})^{B A} F_{j B} h^{j} & &
\overline{h}^{A} = - (F^{-1})^{A B} F_{B \overline{\jmath}} \overline{h}^{
\overline{\jmath}}
\end{eqnarray}

\noindent
which upon replacement into $\hat{\cal L}$ gives for the auxiliary field
part
\be
e^{-1} \hat{\cal L}_{AUX} =
- h^{i} F_{i C}(F^{-1})^{C D} F_{D \overline{\jmath}}
\overline{h}^{\overline{\jmath}}
\ee

The next step is to perform the duality transformation on the kinetic part.
For that \cite{trecep} we write
\be
t^{A} = Re \, t^{A} + i Im \, t^{A}
\ee

\noindent
then make the replacement $\partial_{\mu} Im \, t^{A} \rightarrow L_{\mu}^{A}$
and add the necessary Lagrange multipliers:
\begin{eqnarray}
e^{-1} \hat{\cal L}'_{KIN} & = &
- F_{A B} \partial_{\mu} Re \, t^{A} \partial^{\mu} Re \, t^{B} -
(F_{j A} D^{-}_{\mu} z^{j} +
F_{A \overline{\jmath}}  D^{-}_{\mu} \, \overline{z}^{\overline{\jmath}})
\partial^{\mu} Re \, t^{A} \nonumber \\
 & &- F_{A B} L_{\mu}^{A} L^{\mu B} - 2 F_{A} L_{\mu}^{A} H^{\mu}
- i \, (F_{A \overline{\jmath}} D^{- \mu} \,
\overline{z}^{\overline{\jmath}} - F_{j A} D^{- \mu} z^{j}) L_{\mu}^{A}
\nonumber \\
 & &- \frac{1}{2} e^{-1} \epsilon^{\mu \nu \rho \sigma} b_{\mu \nu A}
\partial_{\rho} L_{\sigma}^{A}
\end{eqnarray}

\noindent
The equation of motion for $L_{\mu}^{A}$ is then

\be
L^{\mu B} = (F^{-1})^{B C} \left[ - F_{C} H^{\mu} + \frac{1}{2} v_{C}^{\mu}
+ \frac{i}{2} \left( F_{j C} D^{- \mu} z^{j} -
F_{C \overline{\jmath}} D^{- \mu} \, \overline{z}^{\overline{\jmath}}
\right) \right]
\ee

\noindent
with
\be
v_{A}^{\mu} = - \frac{1}{2} e^{-1} \epsilon^{\mu \nu \rho \sigma}
\partial_{\nu} b_{\rho \sigma A}
\ee

\noindent
Replacing (4.7) in (4.6) we get finally
\begin{eqnarray}
e^{-1} \hat{\cal L}'_{KIN} & = &
- F_{A B} \partial_{\mu} Re \, t^{A} \partial^{\mu} Re \, t^{B} -
(F_{j A} D^{-}_{\mu} z^{j} +
F_{A \overline{\jmath}}  D^{-}_{\mu} \, \overline{z}^{\overline{\jmath}})
\partial^{\mu} Re \, t^{A} \nonumber \\
 & &+ \frac{1}{4} (F^{-1})^{A B}
\left[ v_{A}^{\mu} - 2 F_{A} H^{\mu} + i \, (F_{j A} D^{- \mu} z^{j} -
F_{A \overline{\jmath}} D^{- \mu} \, \overline{z}^{\overline{\jmath}})
\right] \nonumber \\
 & & \times
\left[ v_{B \mu} - 2 F_{B} H_{\mu} + i \, (F_{j B} D^{-}_{\mu} z^{j} -
F_{B \overline{\jmath}} D^{-}_{\mu} \, \overline{z}^{\overline{\jmath}})
\right]
\end{eqnarray}

This expression for $\hat{\cal L}'_{KIN}$ seems to imply (compare with eq.
(4.1)) that the scalar partner of $b_{\mu \nu A}$ is
\be
C_{A} = F_{A}
\ee

Indeed, the kinetic term for the $b_{\mu \nu A}$ field dictates the presence
of the following scalar kinetic term:
\begin{eqnarray}
\lefteqn{- \frac{1}{4} (F^{-1})^{A B} \partial_{\mu} F_{A}
\partial^{\mu} F_{B} = } \nonumber \\
 & &- F_{A B} \partial_{\mu} Re \, t^{A} \partial^{\mu} Re \, t^{B} -
(F_{j A} D^{- \mu} z^{j} +
F_{A \overline{\jmath}}  D^{- \mu} \, \overline{z}^{\overline{\jmath}})
\partial^{\mu} Re \, t^{A} \nonumber \\
 & &- \frac{1}{4} (F^{-1})^{A B}
(F_{j A} D^{- \mu} z^{j} -
F_{A \overline{\jmath}} D^{- \mu} \, \overline{z}^{\overline{\jmath}})
(F_{j B} D^{-}_{\mu} z^{j} -
F_{B \overline{\jmath}} D^{-}_{\mu} \, \overline{z}^{\overline{\jmath}})
\nonumber \\
 & & + (F^{-1})^{A B} F_{i A} F_{B \overline{\jmath}} D^{-}_{\mu} z^{i}
D^{- \mu} \, \overline{z}^{\overline{\jmath}}
\end{eqnarray}

\noindent
where we have used the chain rule as well as the properties that $F$ must
satisfy \cite{sabi}.

Thus (4.9) provides not only the kinetic term for $C_{A}$ in (4.10) but also
the extra kinetic contribution for the $z^{i}$ fields that, together with
(4.4), is precisely what we need in order to have the
appropriate form of the Lagrangian corresponding to the new function
$\tilde F$ that one obtains after completing the dualization procedure.
That function is
\be
\tilde{F} (L,\ldots) =
F(T(L) + \overline{T}(L), \dots) - L_{A} (T(L) + \overline{T}(L))^{A}
\ee

\noindent
where $T(L)$ in (4.12) is given by the solution of
\be
L_{A} = F_{A} = \frac{\partial F}{\partial T^{A}}
\ee

\noindent
whose first component is (4.10) and whose inverse is
\be
T^{A} + \overline{T}^{A} = - \tilde{F}^{A} =
- \frac{\partial \tilde{F}}{\partial L_{A}}
\ee

In order to verify what we have stated above, we do not need to invert
explicitly
(4.10) (in other words, we do not need to find the particular form of $\tilde
F$)
which may be difficult, but it is sufficient to use the following
identities that always hold
\begin{eqnarray}
\tilde{F}_{j} = F_{j} & &
\tilde{F}_{\overline{\jmath}} = F_{\overline{\jmath}} \nonumber \\
\tilde{F}^{A B} = - (F^{-1})^{A B} & &
F_{A B} = - (\tilde{F}^{-1})_{A B} \nonumber \\
\tilde{F}^{A} \, _{j} = (F^{-1})^{A B} F_{j B} & &
F_{j A} = - (\tilde{F}^{-1})_{A B} \tilde{F}^{B} \, _{j} \nonumber \\
\tilde{F}^{A} \, _{\overline{\jmath}} = (F^{-1})^{A B} F_{B \overline{\jmath}}
& &
F_{A \overline{\jmath}} = - (\tilde{F}^{-1})_{A B}
\tilde{F}^{B} \, _{\overline{\jmath}} \nonumber \\
\tilde{F}_{i \, \overline{\jmath}} = F_{i \, \overline{\jmath}}
- F_{i A} (F^{-1})^{A B} F_{A \overline{\jmath}}& &
F_{i \, \overline{\jmath}} = \tilde{F}_{i \, \overline{\jmath}}
+ \tilde{F}^{A} \, _{i} (\tilde{F}^{-1})_{A B} \tilde{F}^{B} \,
_{\overline{\jmath}}
\end{eqnarray}

It is worth stressing that, even though it may be impossible to obtain
explicitly $\tilde F$ from $F$, or viceversa, the hybrid formulation
(as given by (4.4), (4.9) plus the unmodified part of the Lagrangian) can
always be reached from {\it either} $\tilde F$ or $F$.

Let us also mention that, even though we have not done it explicitly here,
it is a simple matter to include the case when some of the $T^{A}$ multiplets
appear in the function $f_{a b}$, where there is an extra Chern-Simons term
modifying $v_{A}^{\mu}$ in (4.7) and (4.9) \cite{trecep,sabi}.

Next we would like to show how the fact that certain particular couplings
of linear multiplets can not be written in standard supergravity form,
manifests
itself in component language. Clearly, the duality transformation to the
old-minimal set of auxiliary fields can always be performed {\it except} when
the $H_{\mu}H^{\mu}$ term in the Lagrangian (4.1) {\it vanishes}, in which case
we can
not solve for $H_{\mu}$ from its equation of motion. Requiring that the total
coefficient of $H_{\mu}H^{\mu}$ vanishes provides us the following
obstruction condition for the function
$F(L_{A}, Z^{i}, \overline{Z}^{\overline{\jmath}})$ in terms of its
$\theta=0$ component
\be
F^{A B} C_{A} C_{B} + 3 \left[ 1 - (nz)^{j} F_{j} \right] +
6 F^{A}\, _{j} C_{A} (nz)^{j} + 9 F_{i \, \overline{\jmath}} (nz)^{i}
(n\overline{z})^{\overline{\jmath}}
= 0
\ee

There are some important aspects to note in this equation. First, only the
function $F$ appears in it; the superpotential and $f_{a b}$ are not involved.
Second, these exceptional theories satisfying (4.16) constitute a very large
class since, if $F_{part}(C,z,\overline{z})$ is a solution, so it will be
\be
F(C,z,\overline{z}) = F_{part}(C,z,\overline{z}) +
F_{(1)}(C) + F_{(0)}(C)
\ee

\noindent
where $F_{(1)}(C)$ and $F_{(0)}(C)$ are arbitrary homogenous functions of
degrees $1$ and $0$ respectively of the $C_{A}$ fields.

A particular solution to (4.16) involving only linear multiplets is easy to
find
\be
F(C) = \sum_{A} \alpha^{A} \; ln C_{A} \; , \;\;\;\; \sum_{A} \alpha^{A} = -3
\ee

\noindent
Another example is provided by the effective Lagrangian for the superstring
\cite{witten} which corresponds to the function
\be
F(L, S + \overline{S}, Z \overline{Z}) = - 3 \;
ln \frac{(S + \overline{S})^{\frac{1}{3}}}{L}  +
3 \; Z \overline{Z} \; (S + \overline{S})^{\frac{1}{3}}
\ee

\noindent
where the chiral multiplet $Z$ has chiral weight $n = 1/3$.

\vskip 30pt

\noindent
\section{Superspace duality and $H^{(2,2)}$ cohomology.}
\setcounter{equation}{0}

In the previous paragraph we have seen that the duality transformation
which maps the $T^{A}$ chiral multiplets into linear multiplets $L_{A}$
(eq. (3.13)) also gives a coordinate transformation between $Re \, t^{A}$
and the $\theta=0$ components of the linear multiplets

\be
C_{A} = L_{A} \mid_{\theta=0}
\ee

This coordinate transformation has a simple interpretation in the Calabi-Yau
geometry. Indeed, the $b_{\mu \nu A}$ component of $L_{A}$ is an element
of $H^{(2,2)}$ while the metric is an element of $H^{(1,1)}$. It is then
natural
to map the metric $G_{i \, \overline{\jmath}}$ into its dual counterpart which
is
an element of
$H^{(2,2)}$. This map induces a reparametrization of the associated moduli
scalars.

For this purpose let us recall that \cite{dieciocho,ventitres}, if
$G_{A B}$ is the metric for the $(1,1)$ forms moduli space, then

\be
G_{A B} \int J \wedge J \wedge J  = \frac{3}{2}
\int V_{A} \wedge ^{\ast}V_{B}
\ee

\vskip .1in

\noindent
where the K\"{a}hler form $J$ can be expanded as follows

\be
J = \sum_{A} \lambda^{A} V_{A} \;\;\;\;\;\;\;\; V_{A} \in H^{(1,1)}
\ee

\noindent
where the $V_{A}$ form a basis of $H^{(1,1)}$.

We also have

\[\int J \wedge V_{A} \wedge V_{B} = d_{A B C} \lambda^{C} ,\]
\[\int J \wedge J \wedge J = d_{A B C} \lambda^{A} \lambda^{B} \lambda^{C}\]
\be
d_{A B C} \lambda^{B} \lambda^{C} =
\int J \wedge J \wedge V_{A} =
\frac{4}{3} G_{A B} \lambda^{B} \int J \wedge J \wedge J
\ee

\vskip .1in

The moduli coordinates $\lambda^{A}$ are defined through the dual homology
$2$-cycles $\gamma_{A}$ by

\be
\lambda^{A} = \int_{\gamma_{A}} J
\ee

\noindent
and are identified with $(T + \overline{T})^{A}/2$.

It is now natural to introduce a ``dual" basis $\Omega^{A}$
in $H^{(2,2)}$ such that

\be
\int V_{B} \wedge \Omega^{A} = \delta_{B}^{A}
\ee

\noindent
which then implies

\be
\lambda^{A} = \int J \wedge \Omega^{A}
\ee

\vskip .1in
Let us next consider the dual of the K\"{a}hler form

\be
^{\ast}J = \sum_{A=1}^{h_{(1,1)}} \lambda^{A} \;\; ^{\ast}V_{A}
\ee

\vskip .1in
\noindent
which we now expand in the dual basis $\Omega^{A}$ defined by (5.6), (5.7):

\be
^{\ast}J = \sum_{A=1}^{h_{(1,1)}} C_{A} \Omega^{A} \int J \wedge J \wedge J
\ee

\vskip .1in
By comparing eq. (5.8) with (5.9) and using (5.2) we find

\be
^{\ast}V_{A} = \frac{2}{3}
G_{A B} \Omega^{A} \int J \wedge J \wedge J
\ee

\noindent
implying

\be
\frac{2}{3} G_{A B} \lambda^{B} = C_{A}
\ee

\noindent
which is nothing but eq. (3.15).

We then may reinterpret the superspace duality relation (3.15) as the change
of coordinates from the basis $V_{A}$ of $H^{(1,1)}$ to the dual basis
$\Omega^{A}$ of $H^{(2,2)}$ with $\Omega^{A}$ defined through eq. (5.6).
Note that the normalization factor in (5.9) is such that

\be
\sum_{A} C_{A} \lambda^{A} = \frac{1}{2}
\ee

\vskip 40pt

\noindent
\section{Conclusions.}

In this note we have considered the dual form of $N=1$,
$D=10$ supergravity and studied its compactification
on a Calabi-Yau threefold, which is known to be a consistent vacuum of the
dual counterpart.

We have presented an off-shell $N=1$ supergravity version of the resulting
$4D$ effective theory which turns out not to be in standard $N=1$
supergravity form.

Several type of duality transformations have been discussed, in particular
the relation between superspace duality and Poincar\'{e} duality of the
internal manifold.

The extension of this Lagrangian to include
effects of the finite size of the five-brane has not been discussed. The
relation between the perturbative expansion versus a strong coupling regime
of the dual theory is also an important question to be addressed in this
context. In particular the dilaton multiplet, which is now a chiral multiplet,
can have non-linear interactions and the theory may exhibit a discrete
duality symmetry similar to that of the moduli fields in string theory
compactification. There is some evidence that, at least for toroidal
compactifications, this symmetry is $SL(2, \bf Z)$ \cite{doce,trece,dieciseis}.
The study of a coordinate free formulation of the linear multiplet geometry,
which is ``dual" to special geometry [27-30] of heterotic
strings is also an interesting aspect to be investigated [26-30].

\vskip 20pt
\noindent
{\bf Acknowledgements}
\vskip 10pt
\noindent

R. D'Auria would like to thank the Physics Department of the University of
California, Los Angeles, for the kind hospitality extended to him during
the completion of this work.

\noindent

S. Ferrara would like to thank the Institute of
Theoretical Physics, Santa Barbara, for his participation in the
``Non-perturbative String Theory Workshop" where part of this investigation
was performed.

\end{document}